# Band Degeneracy and Mott Transition: Dynamical Mean Field Study.

Henrik Kajueter and Gabriel Kotliar
*Department of Physics, Rutgers University, Piscataway, NJ 08855-0849, USA*

We investigate the Mott transition in infinite dimensions in the orbitally degenerate Hubbard model. We find that the qualitative features of the Mott transition found in the one band model are also present in the orbitally degenerate case. Surprisingly, the quantitative aspects of the transition around density one are not very sensitive to orbital degeneracy, justifying the quantitative success of the one band model which was previously applied to orbitally degenerate systems. We contrast this with quantities that have a sizeable dependence on the orbital degeneracy and comment on the role of the intraatomic exchange J.

## I. INTRODUCTION

The Mott transition in transition metal oxides has received renewed theoretical and experimental attention. On the experimental side, new compounds have been synthesized [1,2] and known compounds such as $V_2O_3$ and $NiSeS$ have been studied. New insights have been obtained from studying the one band Hubbard model in the limit of large lattice coordination [14]. Reviews of the subject have appeared recently [3].

Several quantitative comparisons between the physics of three dimensional transition metal oxides and the one band Hubbard model have been performed by now [4,6] The agreement is surprisingly good, if one considers the fact that in three dimensional materials such as $La_{1-x}Sr_xTiO_3$ and the $d^1$ electron has a quasi-threefold degenerate level while the simplest description of $V_2O_3$ involves one electron in a twofold degenerate level [18] . the doping dependence of the physical parameters like the effective electron mass in $La_xSr_{1-x}TiO_3$ can be explained by the one band model without adjustable parameters after the values of $U$ and $D$ have been extracted from photoemission data. Similarly, the single band Hubbard model can describe the temperature dependence of both, the optical and the dc-conductivity in $V_2O_3$ [6]. The photoemission spectra and the optical conductivity of $Sr_{1-x}CaVO_3$ are also in good agreement with the one band Hubbard model [7].

In this paper we examine the *orbitally degenerate* Hubbard model in the limit of infinite dimensions. This publication has two goals: a) using the projective self consistent method we show that all the qualitative aspects of the Mott transition are independent of the band degeneracy. Furthermore, this method can be used to detect the qualitative trends of the dependence of the critical values of the interaction strenght on the band degeneracy. b) To compare the predictions of the one band and the two band Hubbard model to justify the success of the one band model in explaining experimental data in orbitally degenerate systems. For this purpose we use a recent extensions of the IPT to particle hole asymetric problems [19]. In the range of densities between zero and one this method is very accurate and allow us to carry out a detailed comparaison between the one and the two band Hubbard model. We justify a posteriori the success of the one band model for the description of certain quantities related to the low energy features of the Hubbard model in the regime of densities $0 < n < 1$. We also find that the high energy features of the spectra depend on the band degeneracy, as has been known for some time on the basis of less controlled approximations [15].

We also explore the dependence of the low energy and the high energy features on the intraorbital exchange interaction $J$ which only appears in the orbitally degenerate case.

Our Hamiltonian for $l_b = 2$ bands is given by:

$$H = -\frac{t}{\sqrt{d}} \sum_{<ij>,m\sigma} c^{+}_{im\sigma} c_{jm\sigma} + \frac{U_1}{2} \sum_{i,m\sigma} n_{im\sigma} n_{im\bar{\sigma}} + \frac{U_2}{2} \sum_{i,m\sigma} n_{im\sigma} n_{i\bar{m}\bar{\sigma}} + \frac{U_3}{2} \sum_{i,m\sigma} n_{im\sigma} n_{i\bar{m}\sigma} \quad (1)$$

where $m = 1, 2$ ($\sigma = \pm 1$) denotes the band (spin) index. A bar over $m$ or $\sigma$ means the complementary value, i. e. $\bar{m} = 2$ if $m = 1$ and $\bar{m} = 1$ if $m = 2$ ($\bar{\sigma} = -1$ if $\sigma = 1$ and $\bar{\sigma} = 1$ if $\sigma = -1$).

We introduced three interaction parameters $U_i$. $U_1$ describes the on site interaction between two particles in the same band but with opposite spin. $U_2$ relates to a pair of particles with opposite band and spin index. $U_3$ finally concerns the case of equal spin but opposite band index. $U_3$ is by an intraatomic exchange energy $J$ smaller than $U_2$, i. e. $U_3 = U_2 - J$. Rotational symmetry requires $U_2 = U_1 - 2J$ [15].

We study this model on a Bethe lattice with coordination d in the limit of $d \to \infty$. The band density of states in this case is a semicircle with a full bandwith $2D = 4t$.

The lattice model in infinite dimensions is mapped onto an impurity model together with a self consistency condition [5,3].

The relevant impurity problem describes an impurity $f_{m\sigma}$ with band index $m$ and spin $\sigma$ coupled to a bath of conduction electrons ($c_{km\sigma}$). For the frustrated lattice it is given by:

$$
\begin{aligned}
H_{imp} = \epsilon_f \sum_{m\sigma} f^+_{m\sigma} f_{m\sigma} + \sum_{k,m\sigma} \epsilon_k c^+_{km\sigma} c_{km\sigma} \\
+ \sum_{k,m\sigma} V_k \left( c^+_{km\sigma} f_{m\sigma} + f^+_{m\sigma} c_{km\sigma} \right) + \frac{U_1}{2} \sum_{m\sigma} n^{(f)}_{m\sigma} n^{(f)}_{m\bar{\sigma}} \\
+ \frac{U_2}{2} \sum_{m\sigma} n^{(f)}_{m\sigma} n^{(f)}_{\bar{m}\bar{\sigma}} + \frac{U_3}{2} \sum_{m\sigma} n^{(f)}_{m\sigma} n^{(f)}_{\bar{m}\sigma}
\end{aligned}
\tag{2}
$$

where the hybridization function $\Delta(\omega) := \sum_k \frac{V_k^2}{\omega - \epsilon_k - i\eta}$ has to fulfill the self consistency condition:

$$
\Delta(\omega) = t^2 \, G(\omega) \tag{3}
$$

The fact that the general structure of the model is very similar to the one band case, can be understood on the basis of simple physical arguments along the lines which have already been presented in the one band case [3] and the projective self consistent method. We discuss this in section II. To obtain quantitative results we use a generalization of iterative perturbation theory (IPT) to particle hole asymmetric problems which we introduced earlier [19]. This approximation is more accurate than 10 % in the one band Hubbard model, and captures all the qualitative features. The scheme is based on an interpolating self energy which becomes exact in various limits, namely for low and high frequencies as well as in the atomic and the small $U$ limit. The comparaison between the one band and the two band results based on the IPT is presented in section III.

There have been several recent investigations of the degenerate Hubbard model using a variety of techniques, the Gutzwiller approximation [22], the slave boson technique [21] and the quantum Monte Carlo method [23]. On the qualitative level our results are in agreement with these studies.

## II. THE MOTT TRANSITION IN A DEGENERATE HUBBARD MODEL: PROJECTIVE SELF CONSISTENT ANALYSIS

The degenerate Hubbard model is insulating at integer particle number per lattice site provided that $U$ is sufficiently large. Figure 1 displays the total particle density per lattice site as a function of the chemical potential for $U = 4$ obtained with using the IPT method. There is a Mott transition at $n_{tot} = 1$, 2, and 3. Notice that the gap around the $n_{tot} = 1$ transition is larger than the gap around the $n_{tot} = 2$ transition. In addition, we found the critical value $U_c$ for the instability of the metallic solution is larger for $n_{tot} = 2$ than for $n_{tot} = 1$. The transitions at $n_{tot} = 1$ and $n_{tot} = 3$ are equivalent because of particle hole symmetry.

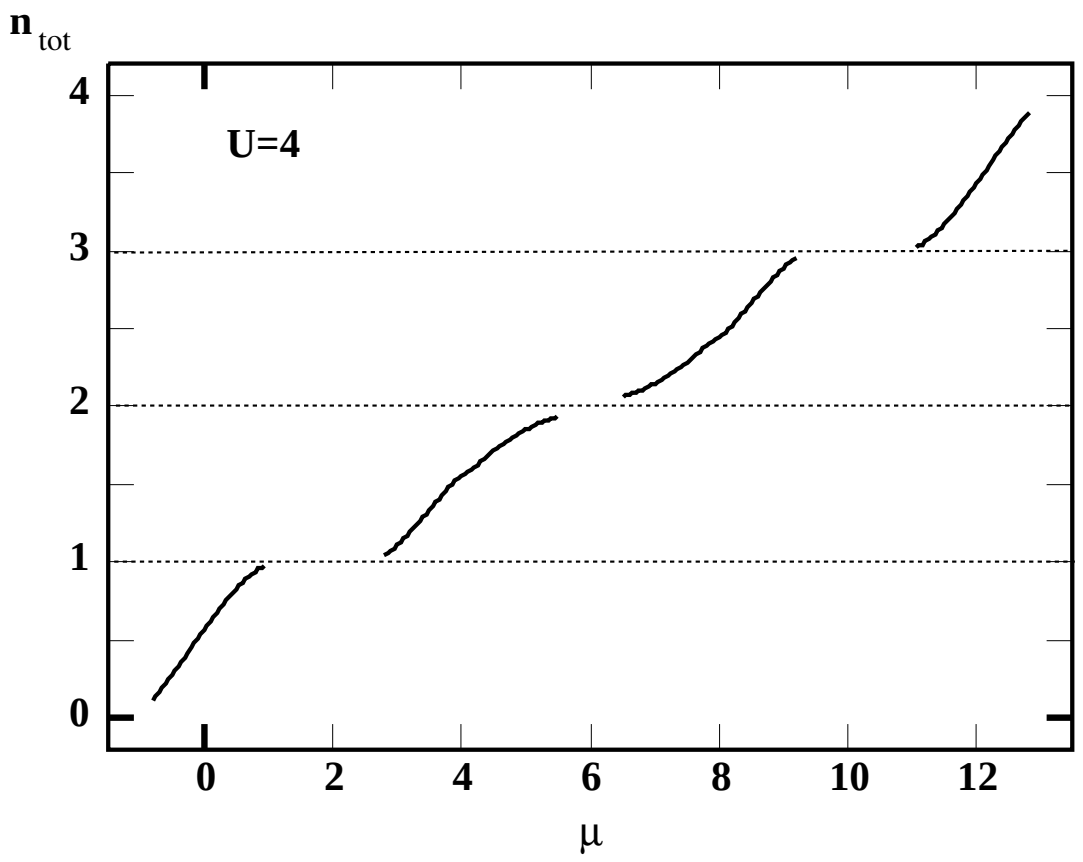

FIG. 1. Particle density of the two band Hubbard model as a function of $\mu$ for $U = 4$.

In the one band Hubbard model, in infinite dimensions, the nature of the Mott transition remained a controversial issue for some time and was finally resolved by the application of the projective self consistent method. [9] [16] We know now that at zero temperature there is a region in the chemical potential and interaction plane where a metallic and an insulating solution coexist. The zero temperature metal to insulator transition is second order and is driven by the divergence of the effective mass.

We have extended the projective self consistent method to the degenerate case and we show that the nature of the Mott transition is not qualitatively changed by orbital degeneracy.

Simple approximations within this scheme explain the differences in the critical $U_c$ and in the size of the band gaps that we observed in IPT and which have also been found in QMC calculations [23].

In the projective self consistent scheme one separates the set $\{k\}$ of "bath" electrons into sets $H \subset \{k\}$ and $L \subset \{k\}$ with corresponding couplings $\{\epsilon_k^H, V_k^H\}$ and $\{\epsilon_k^L, V_k^L\}$ describing the high and low energy features. Then one projects out the high energy degrees of freedom to obtain an effective self consistent problem for the spectral density, involving the low energy variables only.

The weight of the quasiparticle resonance is $w = \sum_{k\in L} \left(V_k^L\right)^2$. For notational clarity we also affix an $L$ to $\epsilon_k$ and $c_{k\alpha}$ for $k \in L$. The quantity $w$ is the small parameter in a singular perturbation theory.

To construct the exact conditions for the Mott transition we separate the impurity Hamiltonian into three parts as

$$\mathcal{H}_{AM} = \mathcal{H}_a + \mathcal{H}_b + \mathcal{H}_m \tag{4}$$

The high energy ("atomic"-like) part $\mathcal{H}_a$ is given by

$$\begin{aligned}\mathcal{H}_a = \frac{U}{2} \sum_{(m\sigma)\neq(m'\sigma')} n_{m\sigma} n_{m'\sigma'} + \sum_{\alpha,k\in H} V_k^H (f_\alpha^+ c_{k\sigma}^H + h.c.) \\ + \sum_{\alpha,k\in H} (\epsilon_k^H - \mu)(c_{k\alpha}^H)^+ c_{k\alpha}^H\end{aligned} \tag{5}$$

$\mathcal{H}_a$ is the Hamiltonian of the Mott insulator with proper occupancy. It has $d$ degenerate ground states carrying a representation of $SU(N)$ ($d = 2l_b$ for $n_{tot} = 1$ and $d = l_b(2l_b - 1)$ for $n_{tot} = 2$; $N = 2l_b$). We label the ground states as $|l>_a$ and define Hubbard operators $X_{ll'} = |l>_a\ _a<l'|$ acting on this low energy manifold.

The low energy spectrum of the impurity Hamiltonian consists of this degenerate manifold combined with excitations of the low energy bath electrons described by the Hamiltonian

$$\mathcal{H}_b = \sum_{\alpha,k\in L} {\epsilon_k}^L ({c^L}_{k\alpha})^\dagger {c^L}_{k\alpha} \tag{6}$$

These low energy states are mixed with the high energy excitations of $\mathcal{H}_a$ by

$$\mathcal{H}_m = \sum_{\alpha,k\in L} V_k^L (f_\alpha^+ c_{k\alpha}^L + hc) = \mathcal{V}^{LH} + \mathcal{V}^{HL} \tag{7}$$

since $f_\alpha^\dagger$ or $f_\alpha$ acting on $|l>_a$ create particle or hole states with energies increased by at least the particle or hole gap. It is convenient to define normalized operators of the local low energy bath electrons at the impurity site,

$$c_\alpha := \frac{1}{\sqrt{w}} \sum_{k\in L} V_k^L c_{k\alpha}^L \tag{8}$$

The total Hamiltonian is then of the form $\mathcal{H}_{AM} = \begin{pmatrix} \mathcal{H}_b & \mathcal{V}^{LH} \\ \mathcal{V}^{HL} & \mathcal{H}_a \end{pmatrix}$. Applying a canonical transformation, one obtains an effective Hamiltonian which is block diagonal and does not mix low and high energy states: $\mathcal{H}_{eff} := e^S \mathcal{H}_{AM} e^{-S} = \begin{pmatrix} \mathcal{H}_{\text{eff}}^L & 0 \\ 0 & \mathcal{H}_{\text{eff}}^H \end{pmatrix}$. Here, the effective Hamiltonian $\mathcal{H}_{\text{eff}}^L$ is defined on the low energy Hilbert space $\{|l>_a\} \otimes \{k \in L\}$. Analogously, the operators $f_\alpha$ are transformed into $F_\alpha := e^S f_\alpha e^{-S} = \begin{pmatrix} F_\alpha^{LL} & F_\alpha^{LH} \\ F_\alpha^{HL} & F_\alpha^{HH} \end{pmatrix}$. In particular, we obtain $F_\alpha^{LL} = \sqrt{w} X_{ll'} J_{ll'}^{\mu\alpha} c_\mu$ where

$$\begin{aligned}J_{ll'}^{\mu\alpha} := \langle l| f_\mu^+ \frac{1}{H - E_{gs}} f_\alpha |l'\rangle \\ - \langle l| f_\alpha^+ \frac{1}{H - E_{gs}} f_\mu |l'\rangle\end{aligned} \tag{9}$$

and summation over repeated indices is assumed

The indices $\mu$, $\mu'$, and $\alpha$ in (9) denote the $2l_b$ spin orbital combinations $(m\sigma)$. The same is true for all other indices as far as $n_{tot}=1$ is concerned. If $n_{tot}=2$ and $l_b=2$, however, the indices $l$, $l'$, $p$, and $p'$ should be thought of as double indices, e.g. $l=(l_1,l_2)$, describing one of six possible pairs $(m\sigma, m'\sigma')$, which denote the quantum numbers of the ground state.

If $n_{tot}=1$ ($l_b=1$ or $l_b=2$), the $J_{ll'}^{\mu\alpha}$ are of the form $J_{ll'}^{\mu\alpha}=a_1\,\delta_{ll'}\delta_{\mu\alpha}+b_1\,\delta_{l'\alpha}\delta_{l\mu}$, while for $n_{tot}=2$ ($l_b=2$) one has $J_{ll'}^{\mu\alpha}=a_2\,\delta_{l_1l'_1}\delta_{l_2l'_2}\delta_{\mu\alpha}-b_2\,\epsilon^{\rho\alpha l_1l_2}\epsilon^{\rho\mu l'_1l'_2}$.

The low energy part of the self consistency condition becomes $t^2G^{LL}(\omega)=\sum_{k\in L}\frac{(V_k^L)^2}{\omega-\epsilon_k^L}$, where $G^{LL}(t-t')=\left\langle TF_\alpha^{LL}(t)\left(F_\alpha^{LL}(t')\right)^+\right\rangle_{\mathcal{H}_{\rm eff}^L}$.

In addition, one can write the effective low energy Hamiltonian as

$$\mathcal{H}_{\rm eff}^L=\mathcal{H}_b+wD^2[J_sO_s+J_pO_p]+\text{const} \tag{10}$$

where $O_s$ and $O_p$ are the $SU(N)$ ($N=2l_b$) generalizations of the spin spin interaction and the potential scattering between the impurity states and the low energy part of the bath of conduction electrons, respectively. The parameters $J_s$, $J_p$ are linear combinations of the exchange matrix elements in eq. 9 and will be discussed below. For $n_{tot}=1$ and $SU(N=2l_b)$, $O_s$ has the form

$$O_s=\left(X_{\mu\mu'}c_\mu c_{\mu'}^+-\frac{1}{N}X_{\mu\mu}c_{\mu'}c_{\mu'}^+\right) \tag{11}$$

which for $l_b=1$ (i. e. $N=2$) can be written as $O_s=4\,\vec{S}c_\sigma^+\vec{\sigma}_{\sigma\sigma'}c_{\sigma'}$. For $l_b=2$ and $n_{tot}=2$, $O_s$ is given by

$$O_s=\left(\frac{1}{2}X_{pp}c_\mu c_\mu^+-\epsilon^{\rho\mu p'_1p'_2}\epsilon^{\rho\mu' p_1p_2}X_{pp'}\,c_\mu c_{\mu'}^+\right) \tag{12}$$

The potential scattering is described by

$$O_p=X_{\mu'\mu'}c_\mu^+c_\mu \tag{13}$$

In equations (11) to (13) as well as below summation over repeated indices is assumed.

The condition for the destruction of the metallic solution follows from the self consistency condition and is given by $t^2\left\langle\{\left(F_\alpha^{LL}\right)^+,F_\alpha^{LL}\}\right\rangle=\sum_{k\in L}\left(V_k^L\right)^2$. Evaluating this expression we obtain:

$$\begin{aligned}1=&\frac{t^2}{2l_b}\left\langle X_{pp'}c_\mu c_{\mu'}^+\right\rangle\left\{J_{pl'}^{\mu\alpha}\left(J_{p'l'}^{\mu'\alpha}\right)^\star-J_{lp'}^{\mu\alpha}\left(J_{lp}^{\mu'\alpha}\right)^\star\right\}\\&+\frac{t^2}{2l_bd}\sum_{ll'\mu\alpha}|J_{ll'}^{\mu\alpha}|^2\end{aligned} \tag{14}$$

It assumes a very simple form in specific cases: For $n_{tot}=1$ and $l_b=1$

$$\frac{1}{t^2}=(\frac{3}{4}-\langle O_s\rangle)\,J_s^2+{J_p}^2 \tag{15}$$

with $J_s=b_1$ and $J_p=a_1+\frac{b_1}{2}$, for $n_{tot}=1$ and $l_b=2$

$$\frac{1}{t^2}=(\frac{15}{16}-\langle O_s\rangle)\,J_s^2+{J_p}^2 \tag{16}$$

with $J_s=b_1$ and $J_p=a_1+b_1/4$, and for $n_{tot}=2$ and $l_b=2$

$$\frac{1}{t^2}=\left(\frac{5}{4}-\langle O_s\rangle\right)\,J_s^2+J_p^2 \tag{17}$$

with $J_s=b_2$ and $J_p=a_2-\frac{b_2}{2}$. Here $\langle\ldots\rangle$ indicates the expectation value with respect to the effective low energy Hamiltonian $\mathcal{H}_{\rm eff}^L$.

An expression for the parameters $a_i$ and $b_i$ (and therefore $J_s$ and $J_p$) can be obtained by an expansion in the hybridization. It turns out that the main differences between the transitions at $n_{tot} = 1$ and $n_{tot} = 2$ appear to leading order in this expansion. To illustrate this point let us evaluate $J_{ll'}^{\mu\alpha}$ for the single band model at $n_{tot} = 1$ and for the two band model at $n_{tot} = 1$ as well as $n_{tot} = 2$ (and $J = 0$). We obtain for $n_{tot} = 1$ ($l_b = 1$ or $l_b = 2$) $a_1 = \frac{1}{\mu - U}$ and $b_1 = \frac{1}{\mu} + \frac{1}{U-\mu}$. For $n_{tot} = 2$ and $l_b = 2$, in contrast, one has $a_2 = \frac{1}{\mu - U}$ and $b_2 = \frac{1}{\mu - U} + \frac{1}{2U-\mu}$. These results can be used to estimate $U_{c1}$, the critical value of the interaction where the insulator ceases to exist, as well as the regime of the stability of the insulating solution in the $\mu - U$ plane. As discussed in reference [16] these quantities are obtained by noticing that for small perturbations $\langle X_{pp'} c_\mu c_{\mu'}^+ \rangle \approx \langle X_{pp'} \rangle \langle c_\mu c_{\mu'}^+ \rangle$. The resulting conditions can be used to determine for given value $U$ the size $g_1$ of the regime where the insulating solution is stable ("insulating gap"). Around particle hole symmetry, i. e. for $n_{tot} = l_b$ this quantity is given by $g_1 = 2\Delta\mu_c$, where an analytic expression can be obtained for $\Delta\mu_{c_1}$:

$$\Delta\mu_{c_1}^2 = \left(\frac{U}{2}\right)^2 + \frac{t^2}{2}\left(1 - \sqrt{1 + 2\,(l_b + 1)\left(\frac{U}{t}\right)^2}\right) \tag{18}$$

We see that the insulating gap in the two band model at $n_{tot} = 2$ is smaller than that of the one band model at $n_{tot} = 1$. In addition, it can be shown by evaluating the equations numerically that the gap for $l_b = 2$ and $n_{tot} = 2$ is also smaller than that for $l_b = 2$ and $n_{tot} = 1$. A plot of the critical $\mu_{c_1}$ as a function of $U$ is shown in figure 2. When comparing the insulating gaps for $n_{tot} = 1$, we find that the values are very similar for $l_b = 1$ and $l_b = 2$.

For the critical interaction at which the insulating gap closes we obtain $U_{c_1}(l_b = 2, n_{tot} = 1) = 1.63 < U_{c_1}(l_b = 1, n_{tot} = 1) = \sqrt{3} < U_{c_1}(l_b = 2, n_{tot} = 2) = \sqrt{5}$.

Notice that $U_{c_1}(l_b = 2, n_{tot} = 1) = 1.63$ is only slightly below $U_{c_1}(l_b = 1, n_{tot} = 1) \approx 1.73$.

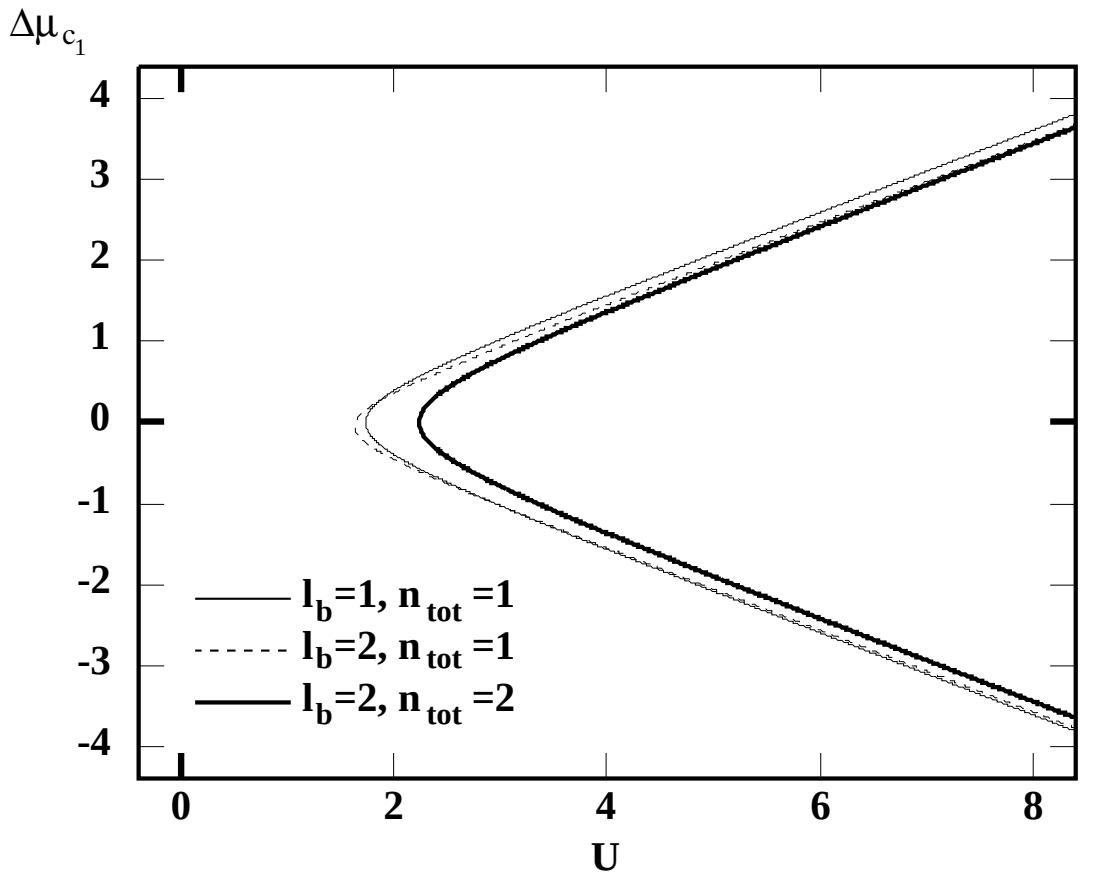

FIG. 2. Stability of the insulating solution: $\Delta\mu_{c_1}$ as a function of $U$ for a different number of bands $l_b$ and fillings $n_{tot} = 1$ or 2. For $n_{tot} = 1$, $\Delta\mu_{c_1} := \mu_{c_1} - \frac{1}{2}U$ while for $n_{tot} = 2$, $\Delta\mu_{c_1} := \mu_{c_1} - \frac{3}{2}U$.

In the large d Hubbard model, the single particle gap or more precisely the jump of the chemical potential when going from hole doping to electron doping is smaller than the optical gap in the insulator [16]. We will denote this gap by $g_2$ and it can be extracted from the $n$ vs $\mu$.

A numerical determination of $g_2$ requires the evaluation of the expectation values $\langle O_s \rangle$. Experience with the one band case has shown that these expectation values are in the intermediate coupling range. A detailed numerical evaluation of these is left for a future publication. It is instructive however to understand the trends that appear as we change band dgeneracy, by taking one half of their values in the strong coupling (sc) limit, an assumption which is very accurate in the one band case [9] [16] . The strong coupling limit can be calculated analtyically $< O_s >_{sc} = -\frac{3}{2}$ for $n_{tot} = 1$ and $l_b = 1$, $< O_s >_{sc} = -5/4$ for $n_{tot} = 1$ and $l_b = 1$, as well as $< O_s >_{sc} = -5$ for $n_{tot} = 2$ and $l_b = 2$. To estimate $g_2$ as well as the critical value $U_{c_2}$ at which the metallic solution breaks down, we assume therefore $< O_s > = \frac{1}{2} < O_s >_{sc}$. This yields $U_{c_2} = 2.4$ for $l_b = 1$ and $n_{tot} = 1$, $U_{c_2} = 2.5$ for $l_b = 2$ and $n_{tot} = 1$, and $U_{c_2} = 3.9$ for $l_b = 2$ and $n_{tot} = 2$. Notice, that the values for $n_{tot} = 1$ are again very similar.To decide which value is smaller in the exact solution, the computation of the first correction in $\frac{t}{U}$ along the same lines as in the one band model [17] are necessary. The critical interaction for $l_b = 2$ and $n_{tot} = 2$, in contrast, is clearly larger than for $n_{tot} = 1$. The region in the $\mu - U$ plane where the metallic solution is stable is indicated in figure 3, where we plotted $\Delta\mu_{c_2}$ as a function of $U$. For $n_{tot} = 1$, this quantitiy is defined by $\Delta\mu_{c_2} := \mu_{c_2} - \frac{1}{2}U$ while for $n_{tot} = 2$, $\Delta\mu_{c_2} := \mu_{c_2} - \frac{3}{2}U$. $g_2$ depends only weakly on orbital degeneracy, for $U = 4$ and $n_{tot} = 1$ we obtain, for example, $g_2 = 2.6$ for $l_b = 1$ and $g_2 = 2.4$ for $l_b = 2$. On the other hand, in the two band model this gap is found to be smaller at $n_{tot} = 2$ than at $n_{tot} = 1$ ($g_2 = 0.7$ for $l_b = 2$, $n_{tot} = 2$, and $U = 4$).

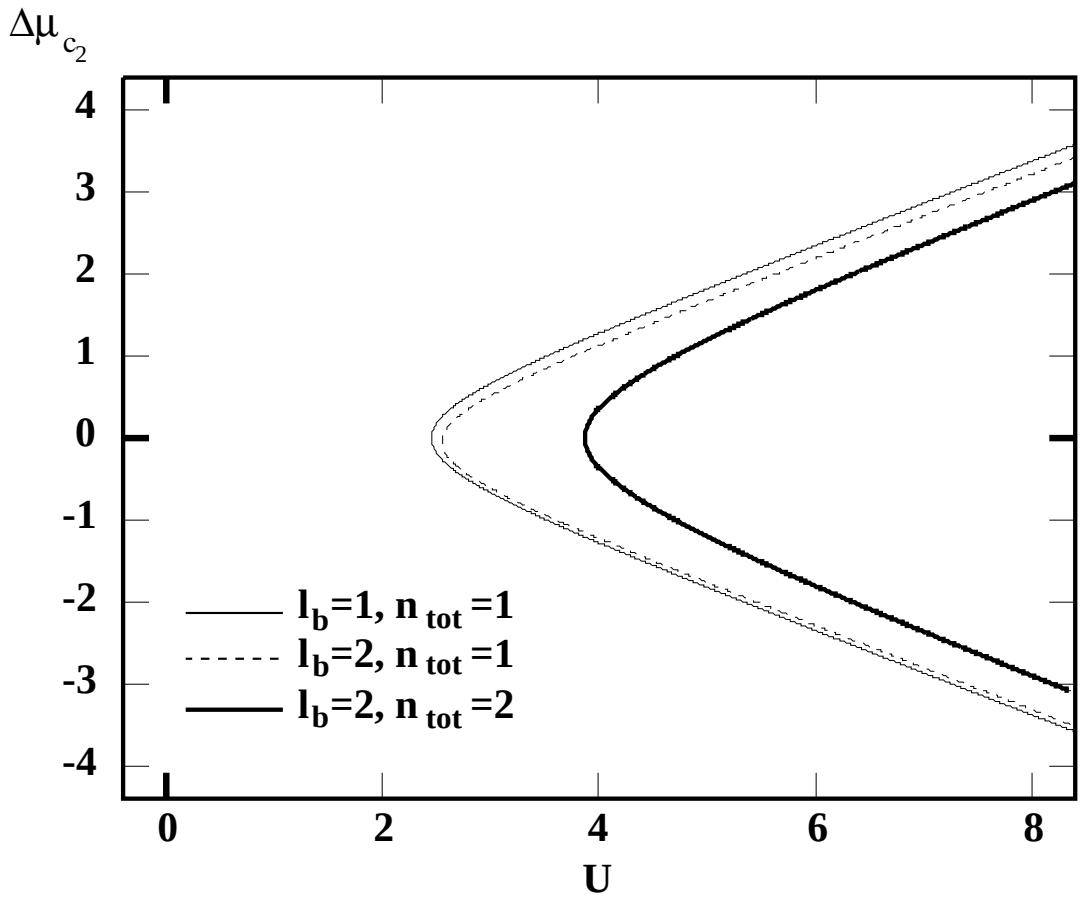

FIG. 3. Stability of the metallic solution: $\Delta\mu_{c_2}$ as a function of $U$ for a different number of bands $l_b$ and fillings $n_{tot} = 1$ or 2. For $n_{tot} = 1$, $\Delta\mu_{c_2} := \mu_{c_2} - \frac{1}{2}U$ while for $n_{tot} = 2$, $\Delta\mu_{c_2} := \mu_{c_2} - \frac{3}{2}U$.

All the qualitative results obtained for $U_{c_i}$ and $g_i$ are in agreement with IPT calculations discussed in the next section and with the recent exact QMC calculations of ref. [23].

## III. IPT: COMPARISON BETWEEN THE ONE BAND AND THE TWO BAND MODEL

The Mott transition as a function of filling (from the metallic side) is driven by the collapse of an energy scale, the renormalized Fermi energy and the consequent divergence of the effective mass. The control parameter is either the interaction $U$ or the chemical potential $\mu$. The goal of this section is to compare the results of the two band against the one band model illustrating that many quantities depend weakly on degeneracy, which justifies the success of the one band model in comparing it with experiments in transition metal oxides. We also notice that the high energy features of the spectral function have a sizeable dependence on band degeneracy in the range of densities between zero and one.

The comparaison is done using the recent extension of the IPT scheme to particle hole asymmetric problems. This extension is very accurate for denisties between 0 and 1.

The evolution of the spectral function of the two band model with increasing filling is displayed in figure 4 for $n_{tot} < 1$. We observe a quasiparticle peak whose width determines the quasiparticle residue $Z$ and scales to zero as the Mott transition is approached.

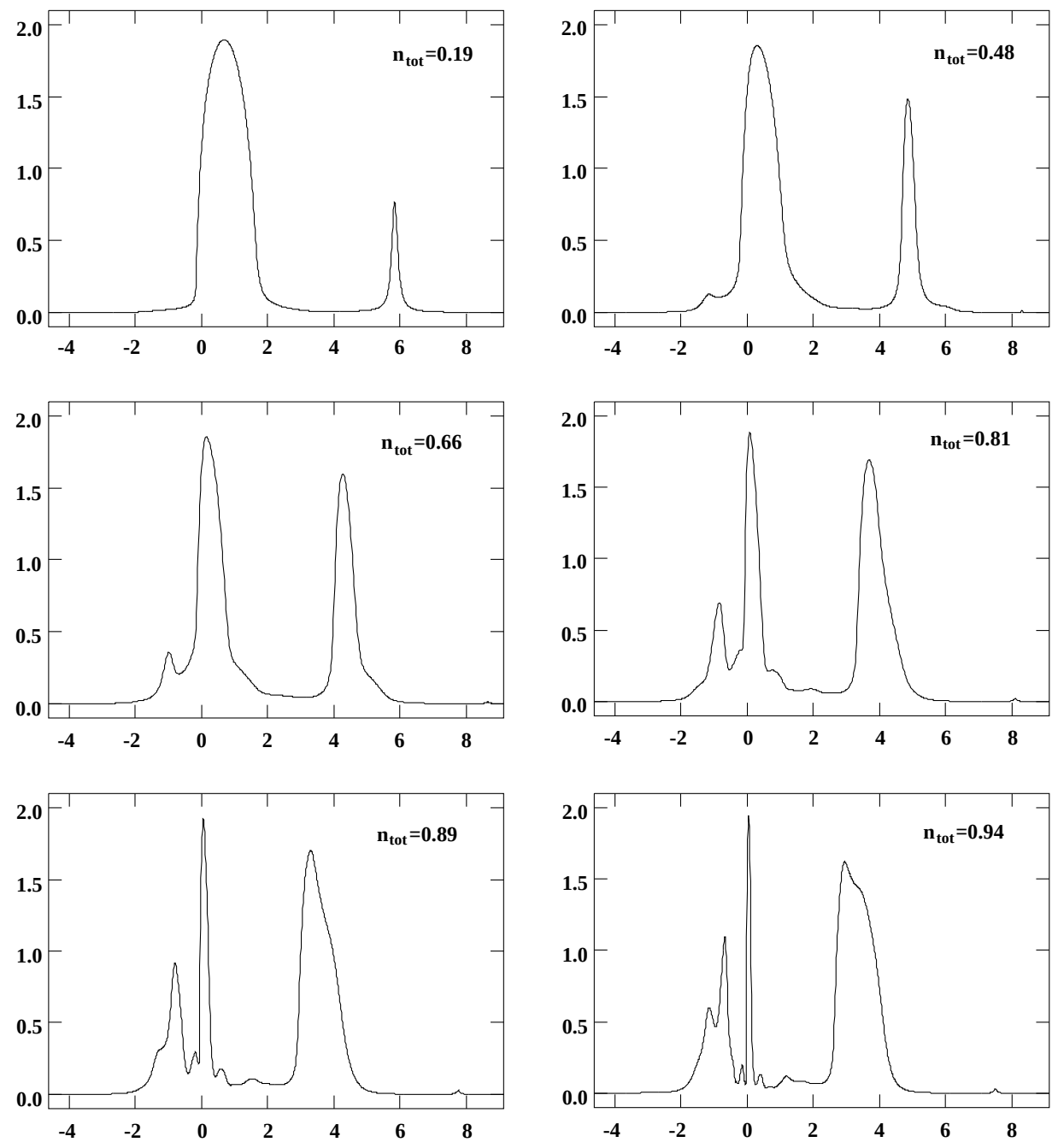

FIG. 4. Evolution of the spectral function of the frustrated two band Hubbard model with increasing filling for $n_{tot} < 1$ and $U = 4$ (J=0).

Figure 5 shows the doping dependence of $Z$. We added the one band result for comparison. The data for the two band case is very close (and slightly smaller) than those for the single band model. The quasiparticle residue vanishes as the Mott transition is approached indicating a divergence of the effective mass.

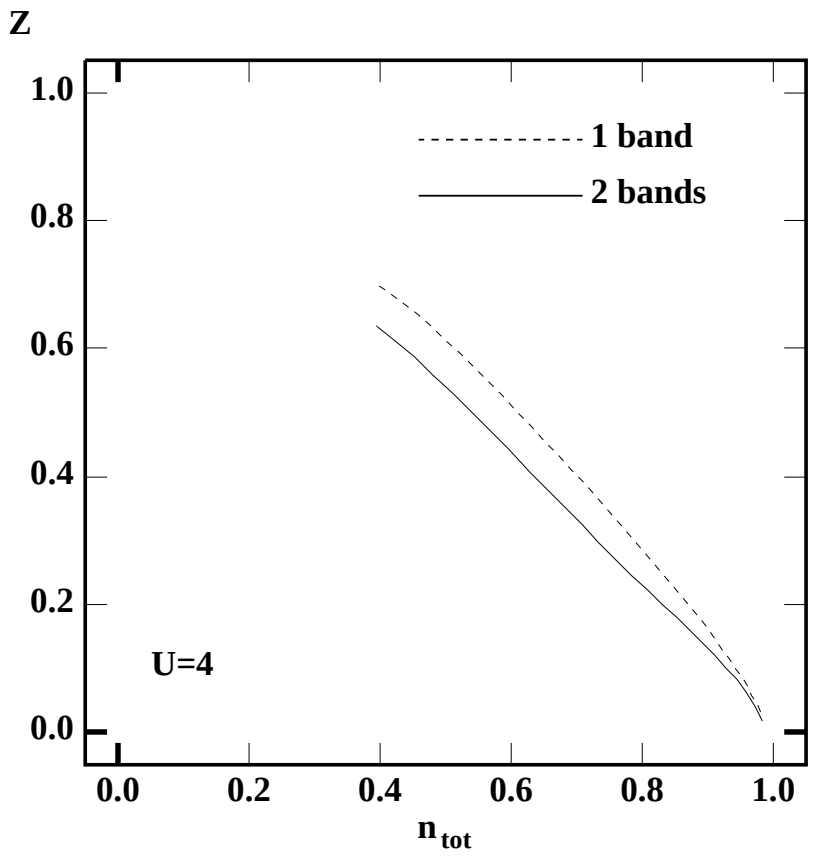

FIG. 5. Quasiparticle residue $Z = \frac{m}{m^{*}}$ as a function of the total particle number $n_{tot}$ for $U = 4$ and $J = 0$.

In figure 6 we show the $U$-dependence of the quasiparticle residue at finite doping. It decreases from its noninteracting value (i. e. $Z(U = 0) = 1$) but remains finite for large $U$.

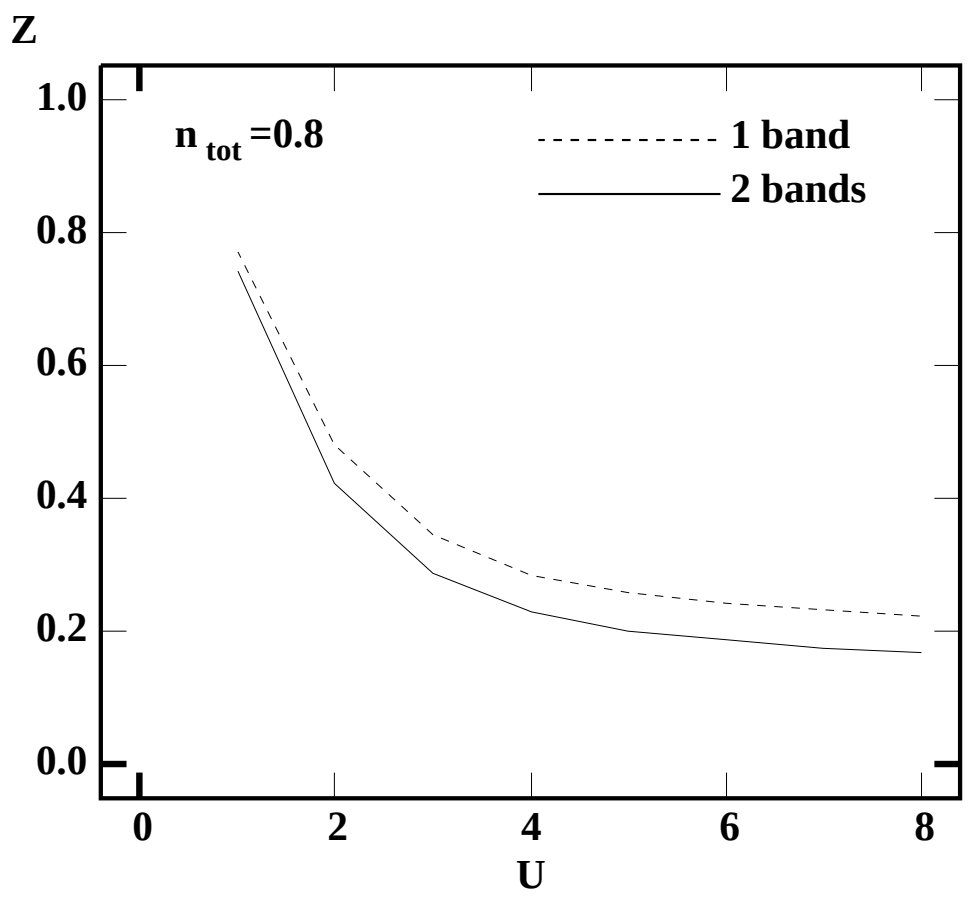

FIG. 6. Dependence of the quasiparticle residue $Z = \frac{m}{m^*}$ on the interaction $U$ for $n_{tot} = 0.8$ and $J = 0$.

At $n_{tot} = 1$, in contrast, $Z$ vanishes above $U_c$ (see figure 7).

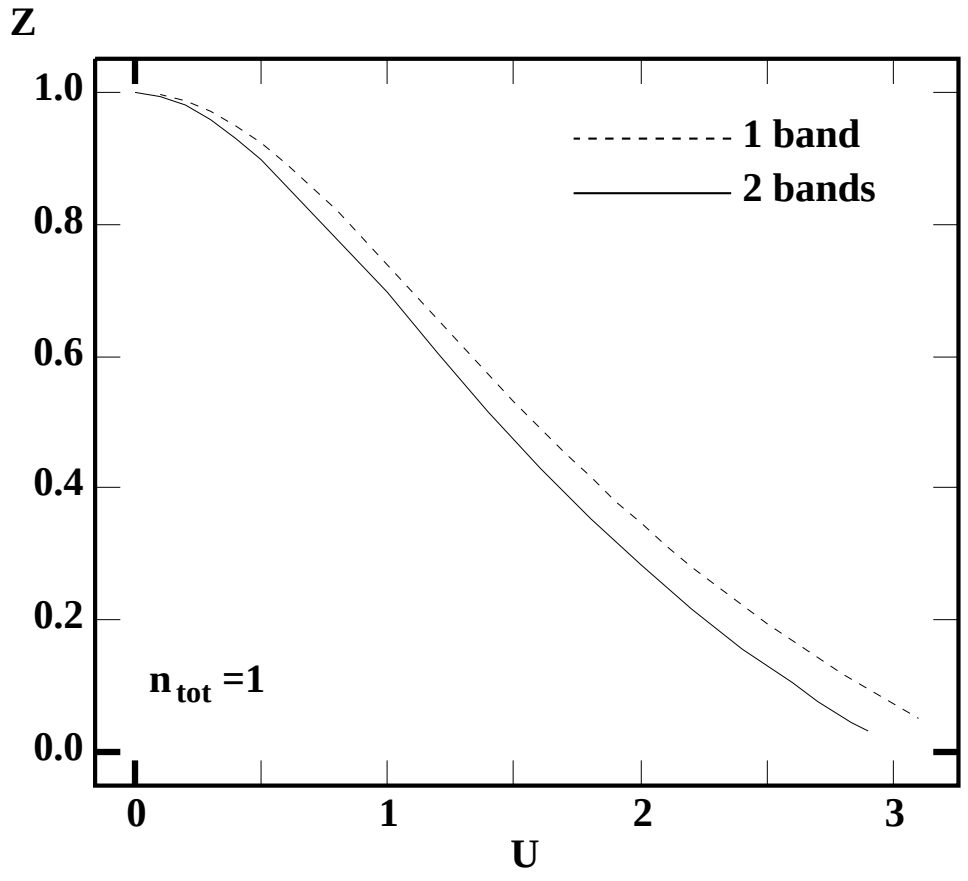

FIG. 7. Comparison of the $Z$ vs $U$ plots for the two band and the single band model at $n_{tot} = 1$ ($J = 0$).

In comparison with the single band model, the slope of the $n$ vs $\mu$ plots is steeper for the case of two bands (figure 8). Thus the compressibility $\kappa = \frac{1}{n^2}\frac{\partial n}{\partial \mu}$ is increased as a result of orbital degeneracy.

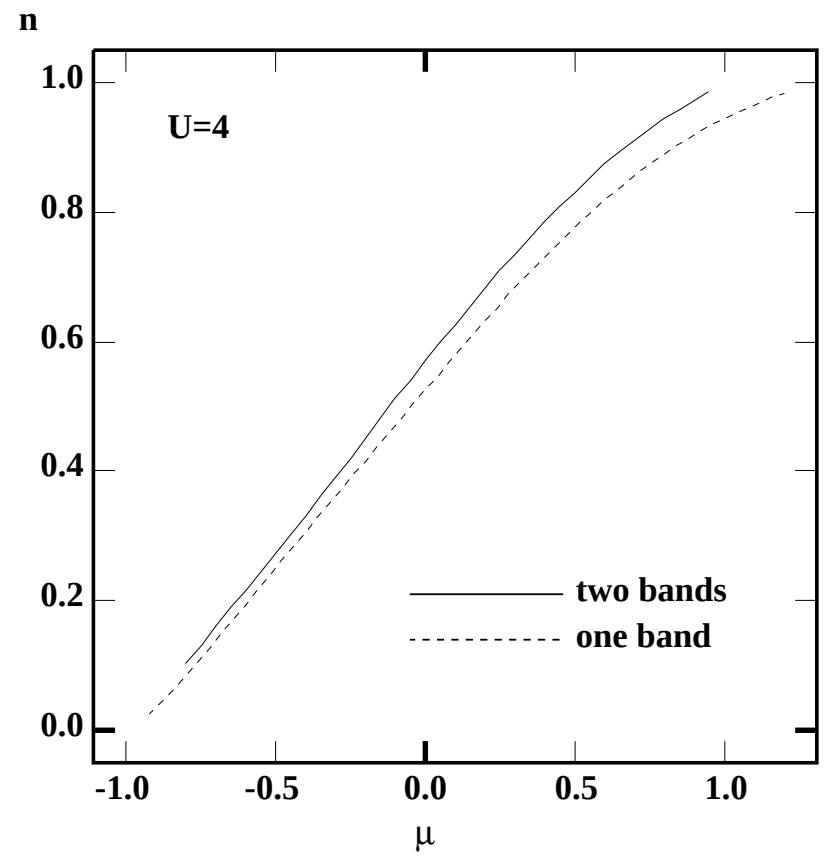

FIG. 8. Comparison of the $n$ vs $\mu$ plots for the two band and the single band model ($U = 4$, $J = 0$).

It is clear that the low energy parameters are not a strong function of the degeneracy as is seen, e. g., from the similarity in the $Z$ vs $U$ curves shown in figures 6 and 7. This is different from the high energy features. Notice the differences in the weights of the Hubbard bands (figure 9) which should be readily observed in photoemission experiments. At $n_{tot} = 1$ the weights of both Hubbard bands are equal (i. e. $\frac{1}{2}$) in the single band model, while in the two band problem the weights are $\frac{1}{4}$ and $\frac{3}{4}$ respectively.

a)

b)

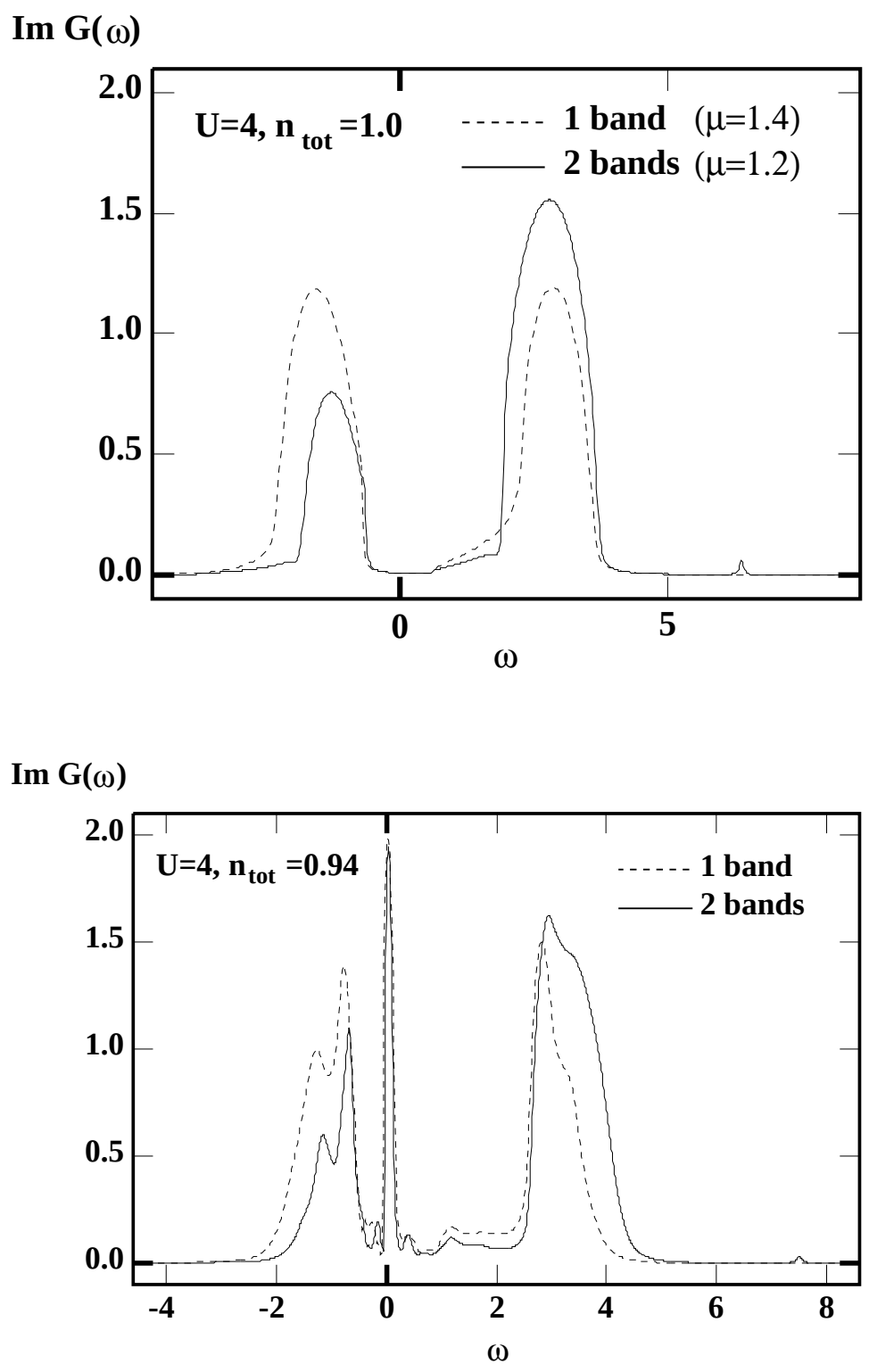

FIG. 9. Comparison of the spectral functions in the one and the two band model for $U = 4$: a) Mott insulator ($n_{tot} = 1$). b) metallic state for small doping ($n_{tot} = 0.94$). The chemical potential for the Mott insulator in figure a) is chosen such that the difference $\mu_{1band} - \mu_{2band}$ is the same as for the metal plotted in figure b.

A more precise probe is given by the optical spectroscopy as discussed in the following section.

## IV. OPTICAL CONDUCTIVITY

Figure 10 contains the optical conductivity $\sigma(\omega)$ corresponding to the plots of figure 9. The data were calculated from:

$$\sigma(\omega) = l_b \sigma_o \, \frac{D^2}{\omega} \int_{-\infty}^{+\infty} d\epsilon \, \rho_o(\epsilon) \int_{-\infty}^{+\infty} \frac{d\omega'}{2\pi} A_\epsilon(\omega') \\ A_\epsilon(\omega' + \omega)(n_f(\omega') - n_f(\omega' + \omega)) \tag{19}$$

where $\sigma_o = \frac{e^2 a^2}{4\nu\hbar}$ and $A_{\epsilon_k}(\omega) = -2\,\mathrm{Im}\, G_k(\omega) = -2\,\mathrm{Im}\frac{1}{\omega+\mu-\epsilon_k-\Sigma_{int}(\omega)}$. The parameter $l_b$ denotes again the number of bands ($l_b = 1$ for one band and $l_b = 2$ for the two band model).

For the insulator (figure 10 a) we find a single maximum around $\omega = U$. The optical conductivity of the metal at small doping (figure 10 b), in contrast, is more complicated. In both, the one and the two band case, there are three main maxima $P_1$, $P_2$, and $P_3$. $P_1$ stems from excitations from the lower Hubbard band to the resonance peak, $P_2$ represents excitations from the resonance peak to the upper Hubbard band, and $P_3$ corresponds to excitations from the lower to the upper Hubbard band. This peak is split into two maxima due to the satellite peaks in the lower Hubbard band. The smaller feature between the Drude peak and $P_1$ for $l_b = 2$ is due to the satellite peaks which appear in the spectral function next to the resonance. In the two band model these satellite peaks are quite pronounced. Notice that $P_1$ has almost the same weight in the one and the two band model because the factor 2 for the orbital degeneracy compensates for the fact that the weight of the lower Hubbard band is reduced by approximately $\frac{1}{2}$. At sufficiently high temperatures all the fine structure disappears and our curves for the one band model are very close to the previous Montecarlo studies by Jarrell et. al. [20].

a)

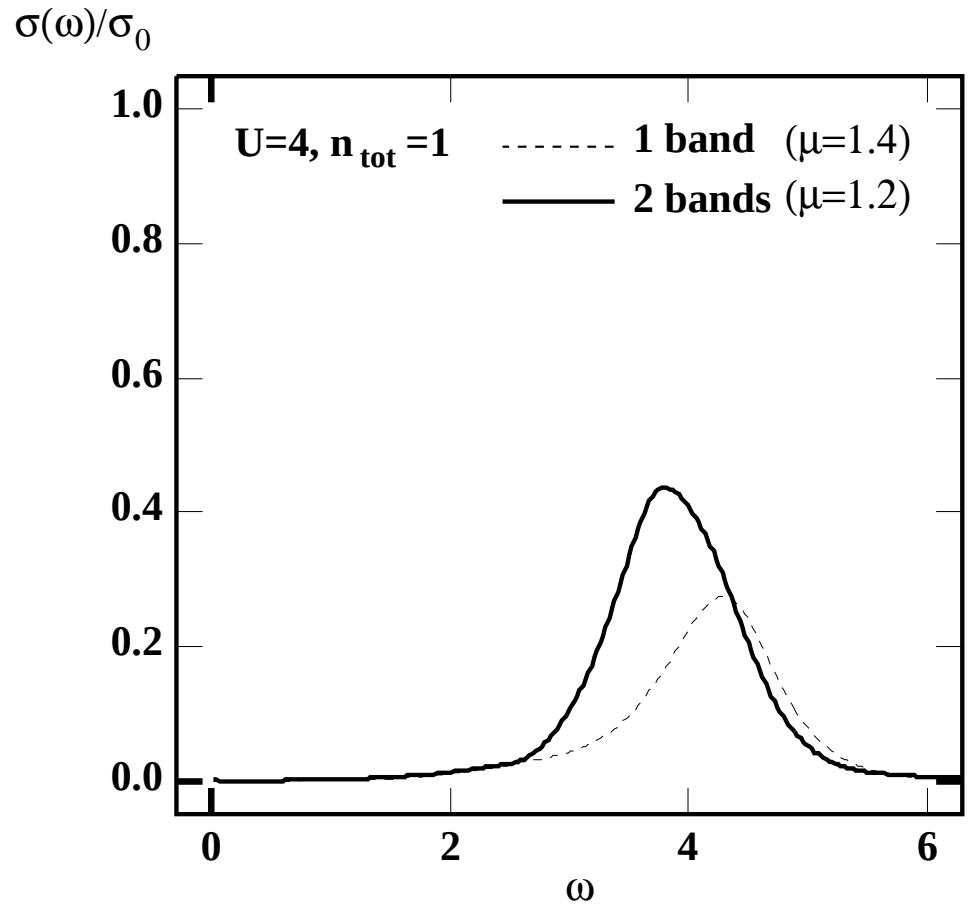

b)

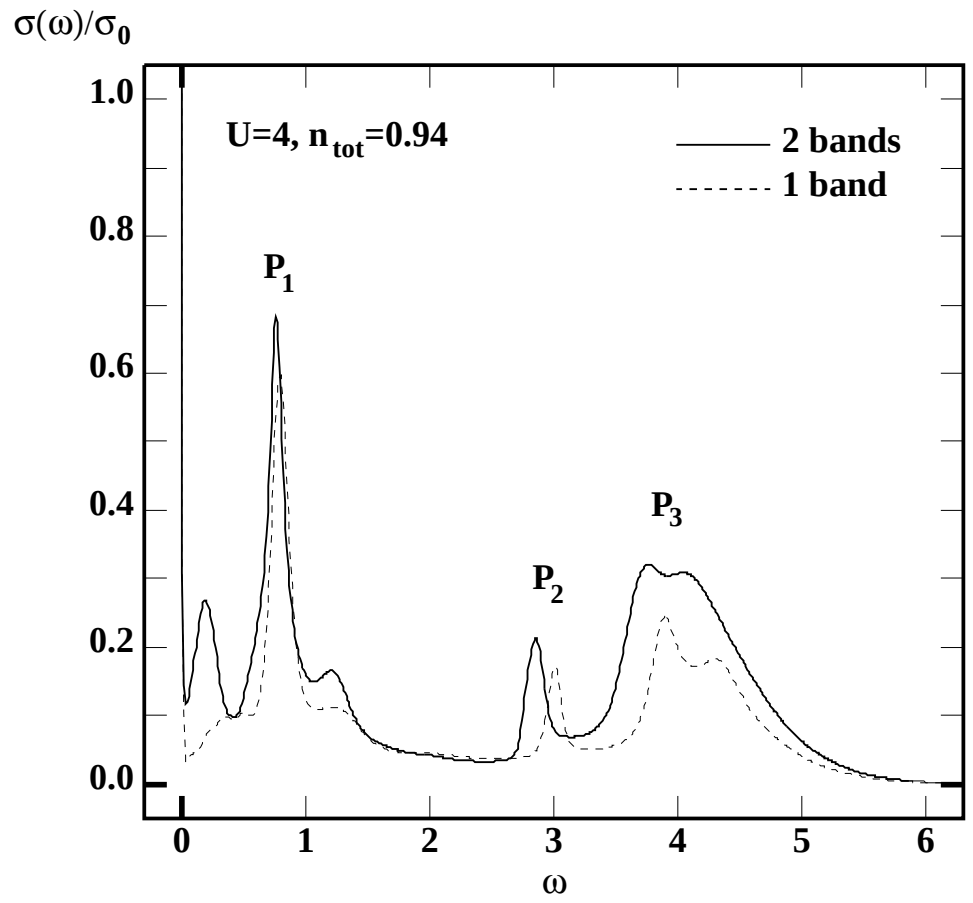

FIG. 10. Comparison of the optical conductivity of the one and the two band model for $U = 4$ (and $J = 0$): a) Mott insulator ($n_{tot} = 1$) and b) metal at small doping ($n_{tot} = 0.94$).

In figure 11 we compare the optical conductivity of the one and two band Hubbard model with one electron per site at $U = 2$. The intensity of the mid infrared peak differs since it originates from both, excitations between the resonance and the lower as well as the upper Hubbard band. The qualitative features however are not changed by orbital degeneracy.

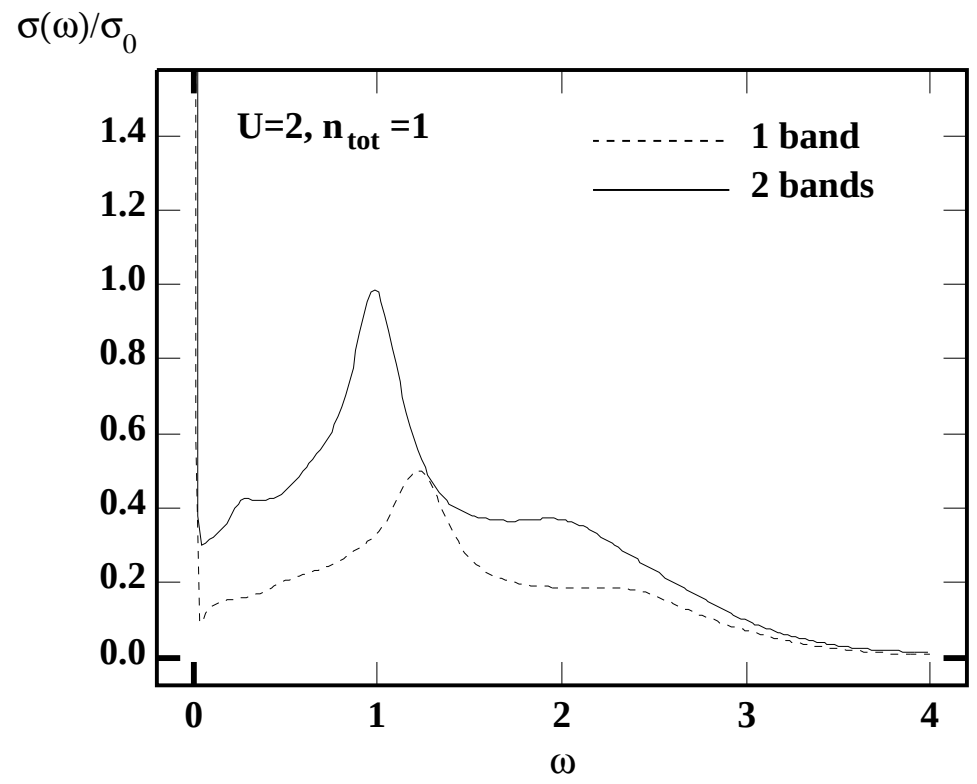

FIG. 11. Comparison of the optical conductivity for the one and the two band model at $U = 2$ and $n_{tot} = 1$.

The optical conductivity of the model, at $T = 0$ can be represented as:

$$\sigma(\omega) = \frac{{\omega_P^*}^2}{4\pi}\delta(\omega) + \sigma_{reg}(\omega) \tag{20}$$

where the coefficient in front of the delta function is the Drude weight and $\omega_P^*$ is the renormalized plasma frequency. The Drude weight and the integrated optical spectral weight evolve as a function of doping. In infinite dimensions the Drude weight is simply related to Z [24,25] via: $\frac{{\omega_P^*}^2}{4\pi} = \frac{4\pi t^2 e^2 a^2 l_b}{\hbar^2 \nu} Z \rho^o(0)$ The integrated optical weight is proportional to the kinetic energy $< T >$. We evaluated the contribution from $\sigma_{reg}(\omega)$ for the cases shown in figures 10 b) and 11. We find that the total spectral weight at finite frequencies $w_{reg} := \int_0^\infty \sigma_{reg}(\omega) d\omega$ is larger for the two band model than in the single band case: $\frac{w_{reg}(l_b=2)}{w_{reg}(l_b=1)} \approx 1.45$ for $U = 4$ and $n_{tot} = 0.94$ (figure 10 b) and $\frac{w_{reg}(l_b=2)}{w_{reg}(l_b=1)} \approx 1.84$ for $U = 2$ and $n_{tot} = 1$ (figure 11.

The evolution of the optical conductivity of the doped metal ($U = 4$) with increasing filling is displayed in figure 12. The data shown correspond to the spectral functions of figure 4. For the dilute system $\sigma(\omega)$ does not have much structure. It increases for $\omega \to 0$ and has a small maximum at high frequencies corresponding to excitations into the upper Hubbard band. As the filling is increased, the resonance peak develops in the density of states. This results in the formation of the associated peaks $P_1$ and $P_2$, indicating excitations between the resonance and the Hubbard bands. At the same time the weight in the upper Hubbard band increases.

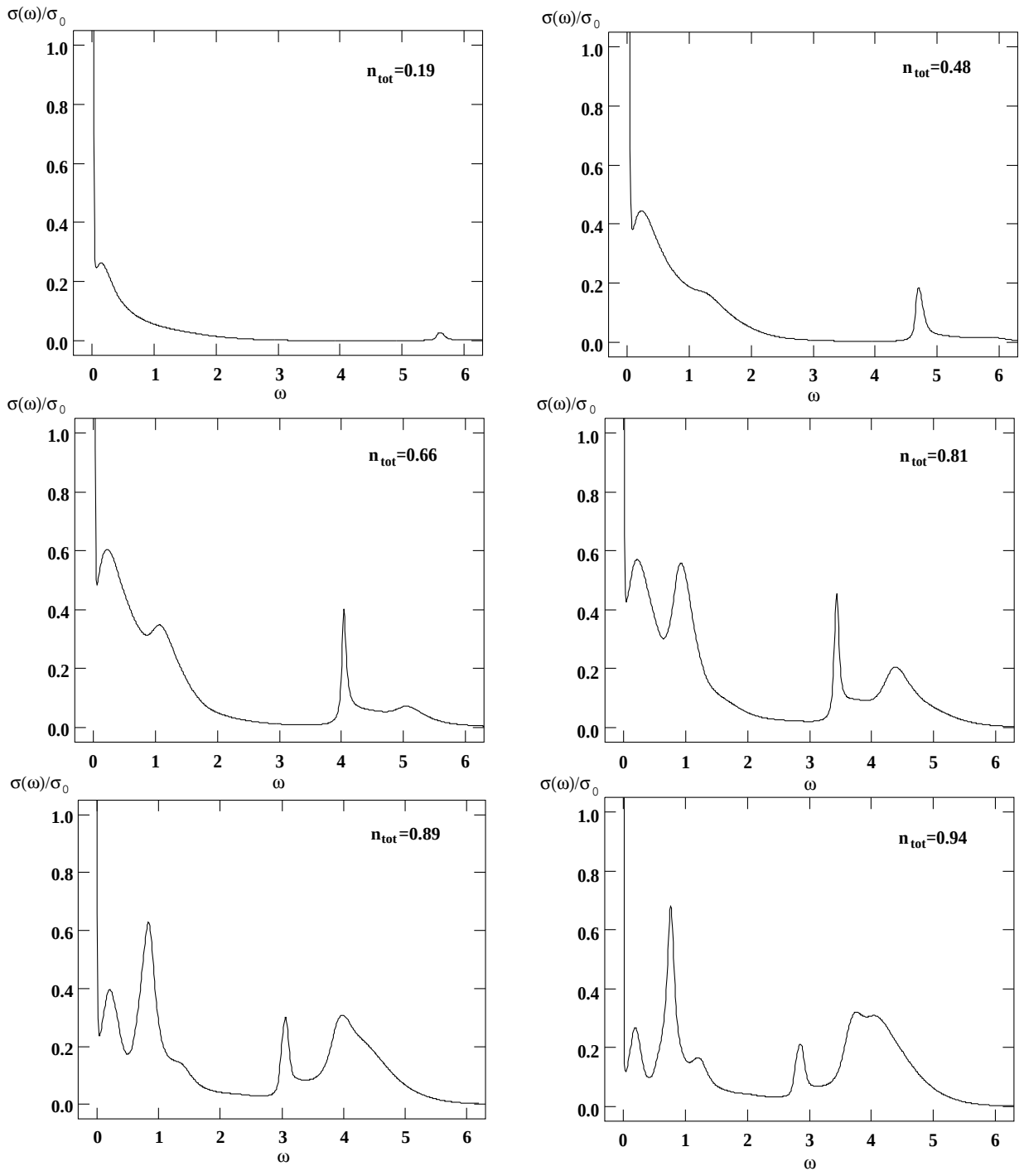

FIG. 12. Evolution of the optical conductivity with increasing filling ($U = 4$, $J = 0$, $l_b = 2$).

Katsufuji *et al.* [12] obtained similar plots for a whole series of substances $R_{1-x}Ca_xTiO_3$ with rare earth ions $R = La$, $Pr$, $Nd$, $Sm$, and $Y$. They observed that an isosbectic point $\omega_c$ where the optical conductivity vs doping curves intersect, and computed tn effective number of carriers:

$$N_{eff} := \frac{2m_0\nu}{\pi e^2} \int_0^{\omega_c} \sigma(\omega) d\omega \tag{21}$$

Fitting a linear behavior to the experimental data for small doping, they showed that the corresponding slope $C$ increases drastically as $U_c$ is approached: $C \sim 1/(U - U_c)$.

Motivated by these experiments we computed the same quantitiy in the degenerate Hubbard model. for $U = 2.5$, 3, and 4 (figure 13). In all three cases the curves are not really linear. The mean slope is enhanced near $U_c = 2.9$. However, comparing the data for $U = 4$ and $U = 3$, we see that the slope does change very much as $U_c$ is approached. These findings are in agreement with Monte Carlo data for the two dimensional Hubbard model [13].

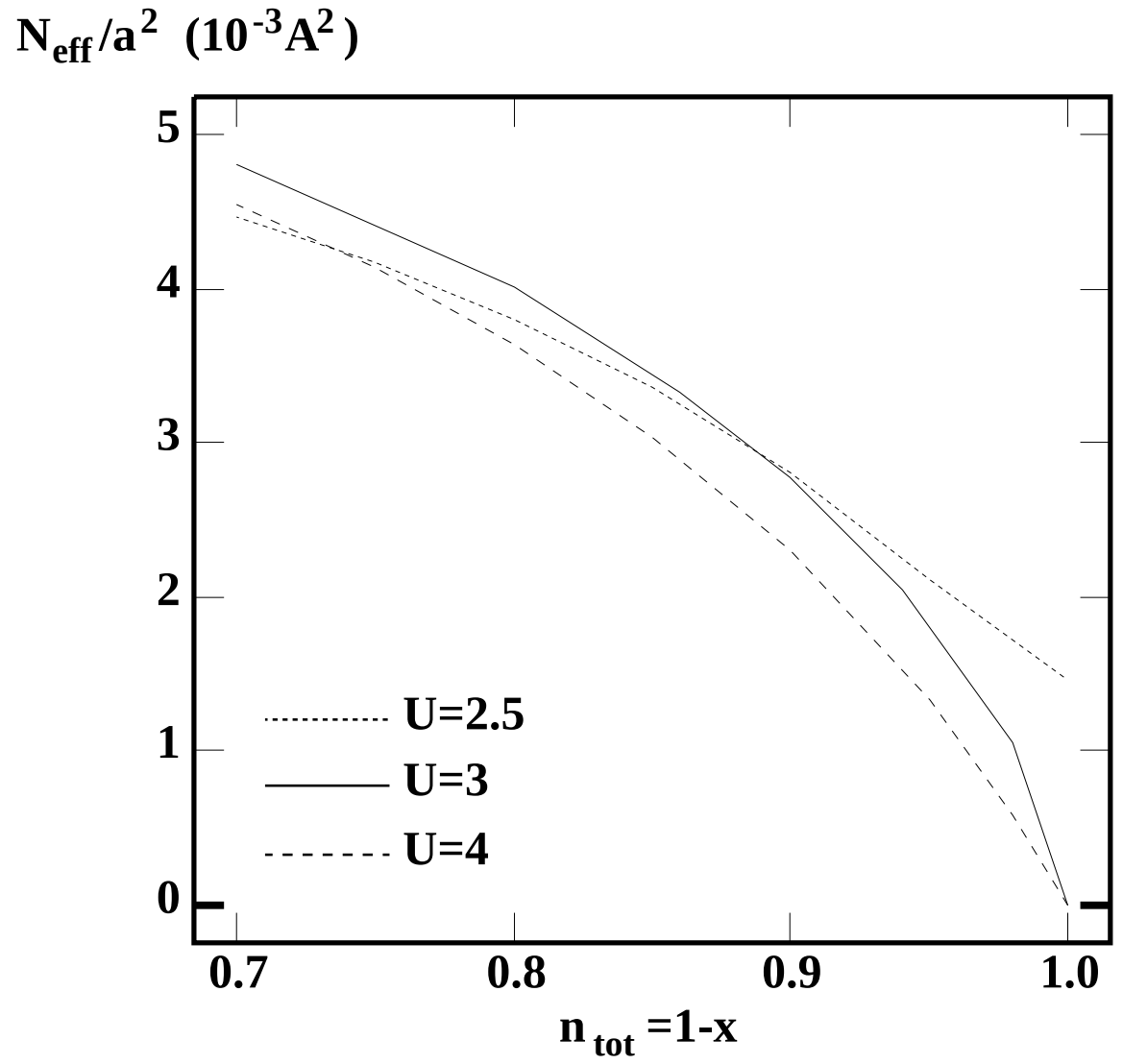

FIG. 13. $N_{eff}/a^2$ for different $U$. The calculations were done at $T = 0.02$ with a finite $\Gamma = 0.2$.

## V. EFFECTS OF A FINITE INTRAATOMIC EXCHANGE $J$

An important parameter with no analogue in the one band model is the intraorbital exchange J, which originates in Hund's rule. This parameter does not affect the low energy features significantly but affects the high energy properties as shown in figure 14. Once $J$ is finite, one is dealing with three different interaction constants ($U_1 = U$, $U_2 = U - 2J$ and $U_3 = U - 3J$) causing a splitting of the upper Hubbard band.

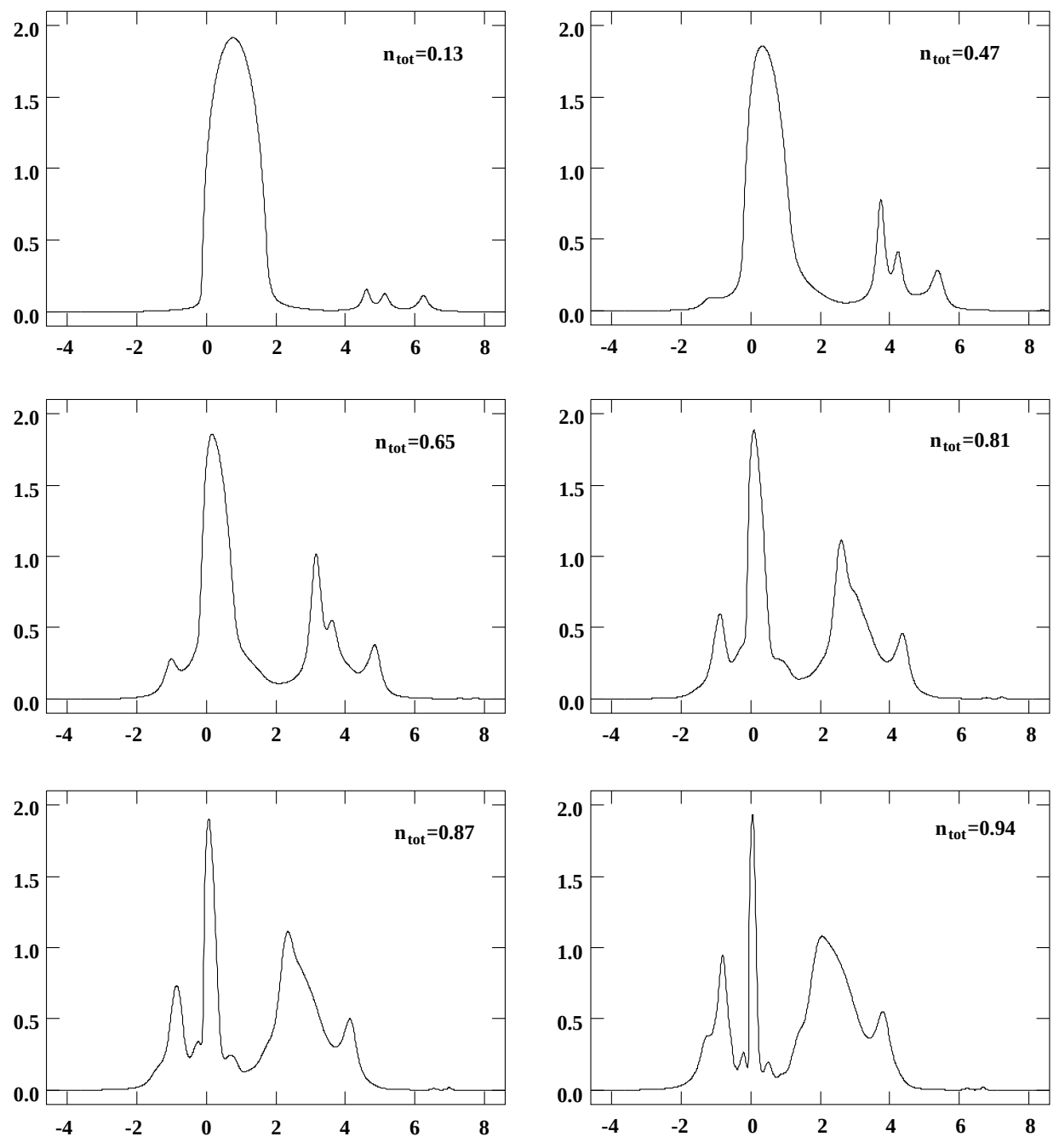

FIG. 14. Evolution of the spectral function for $U = 4$ and $J = 0.5$ with increasing particle density $n_{tot}$.

The interatomic exchange has a profound influence on the phase diagram. A detailed calculation of the competition between the different possible phases depends very sensitively on the nature of the lattice and will not be undertaken here. Instead we will expand on an earlier observation [19] that the one of the main roles of the orbital degeneracy in the context of the dynamical mean field theory is to stabilize over a wide temperature range the paramagnetic solution. In the two band Hubbard model the energy difference between the the ferromagnetic state and the Neel state is given by the energy scale $\frac{t^2 J}{U^2}$, which can be rather small since J tends to be much smaller than U. Since the difference between the various ordered configurations is smaller in the orbitally degenerate situation the system remains in the paramagnetic phase over a wider range of temeperatures.

We present below a simple strong coupling argument as to why this this effect is more pronounced in realistic models which incorporate the hopping of the transition metal ion via the ligand hole. A more quantitative study of the ordered phases of the $Y_{1-x}La_xTiO_3$ system has been carried out recently by Mizoaka and Fujimori using an expansion around the unrestricted Hartree Fock approximation. [26]

We start with a model of $LaTiO_3$ containing three Ti $3d$ orbitals degenerate orbitals ($xy$, $xz$, $yz$). hopping via a filled oxygen shell at an energy $(\epsilon_d - \Delta)$ . The complete Hamiltonian is the sum of three terms:

$$H_{site} = \epsilon_d \sum_{i,\sigma,m} d^+_{im\sigma} d_{im\sigma} + (\epsilon_d - \Delta) \sum_{i,\sigma,l,\delta} p^+_{i+\delta,l,\sigma} \, p_{i+\delta,l\sigma} \qquad (22)$$

Here, $i$ labels the position of the Ti sites. The index $\delta$ is half a unit vector $(\delta \in \{\frac{1}{2}e_x, \frac{1}{2}e_y, \frac{1}{2}e_z\})$ so that $i + \delta$ marks the position of the oxygens. Moreover, $m$ and $l$ are band indexes: $m \in \{xy, xz, yz\}$ and $l \in \{x, y, z\}$.

There are only two pairs of orbitals which overlap: the Ti $3d_{xy}$ with the O $2p_y$ and the Ti $3d_{xz}$ with the O $2p_z$ (xz plane). This results in the kinetic energy:

$$\begin{aligned} H_{kin} = t_{pd} \sum_{i,\sigma;\gamma\in\{-1,1\}} & (d^+_{i,xy,\sigma} \; p_{i+\frac{\gamma}{2}e_x,y,\sigma} + h.c. + d^+_{i,xy,\sigma} \; p_{i+\frac{\gamma}{2}e_y,x,\sigma} + h.c. + \\ & d^+_{i,xz,\sigma} \; p_{i+\frac{\gamma}{2}e_x,z,\sigma} + h.c. + d^+_{i,xz,\sigma} \; p_{i+\frac{\gamma}{2}e_z,x,\sigma} + h.c. + \\ & d^+_{i,yz,\sigma} \; p_{i+\frac{\gamma}{2}e_y,z,\sigma} + h.c. + d^+_{i,yz,\sigma} \; p_{i+\frac{\gamma}{2}e_z,y,\sigma} + h.c.) \end{aligned} \qquad (23)$$

We model the on site interactions by a degenerate Hubbard model:

$$H_{int} = U_1 \sum_{i,m} n_{im\uparrow} n_{im\downarrow} + U_2 \sum_{i,m\neq m'} n_{im\uparrow} n_{im'\downarrow} + U_3 \sum_{i,m\neq m',\sigma} n_{im\sigma} n_{im'\sigma} \qquad (24)$$

Here, $U_1$ is the on site interaction at a Ti site between electrons in the same band and opposite spins. Analogously, $U_2$ ($U_3$) couples electrons in different bands and with opposite (parallel) spins. The parameter $U_3$ is by an intraatomic exchange energy $J$ smaller than $U_2$ ($U_3 = U_2 - J, J > 0$). Moreover, rotational symmetry requires $U_1 - U_2 = 2J$. We neglected the on site interaction on oxygen sites as well as next neighbor interactions. The sums over the spin and band indices are to be understood such that each pair of states $(m\sigma, m'\sigma')$ or $(l\sigma, l'\sigma')$ is included only once.

$$H \;=\; H_{site} \;+\; H_{kin} + H_{int} \qquad (25)$$

We derived the effective Hamiltonian describing the spin excitations of the insulator by using perturbation theory up to fourth order in the O-Ti hopping $t_{pd}$. The procedure is tedious but straigthforward and is most easily summarized by giving the energies of the different magnetic configurations of a pair of neighboring Ti atoms as listed in figure 15.

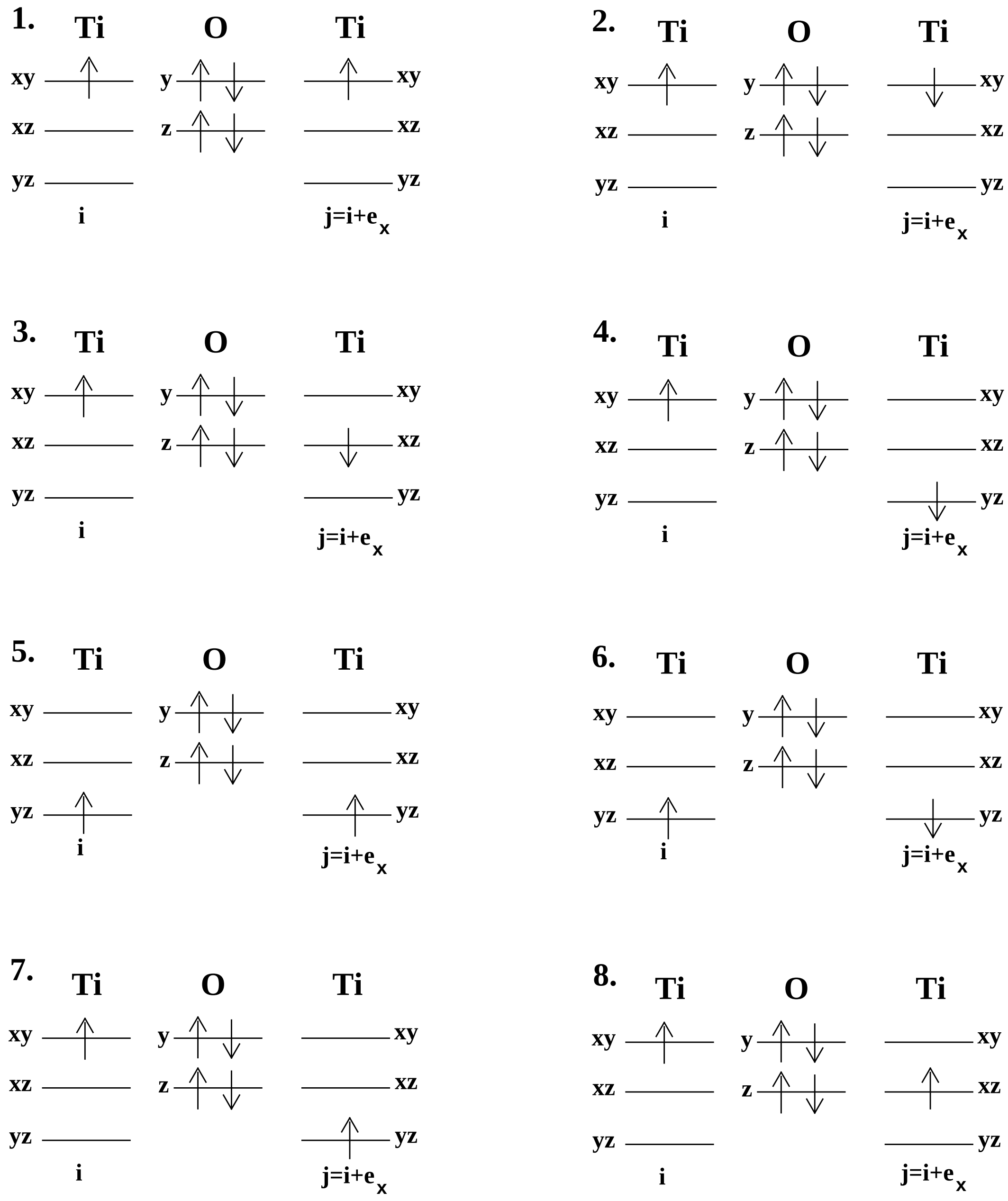

FIG. 15. Independent configurations yielding different energies.

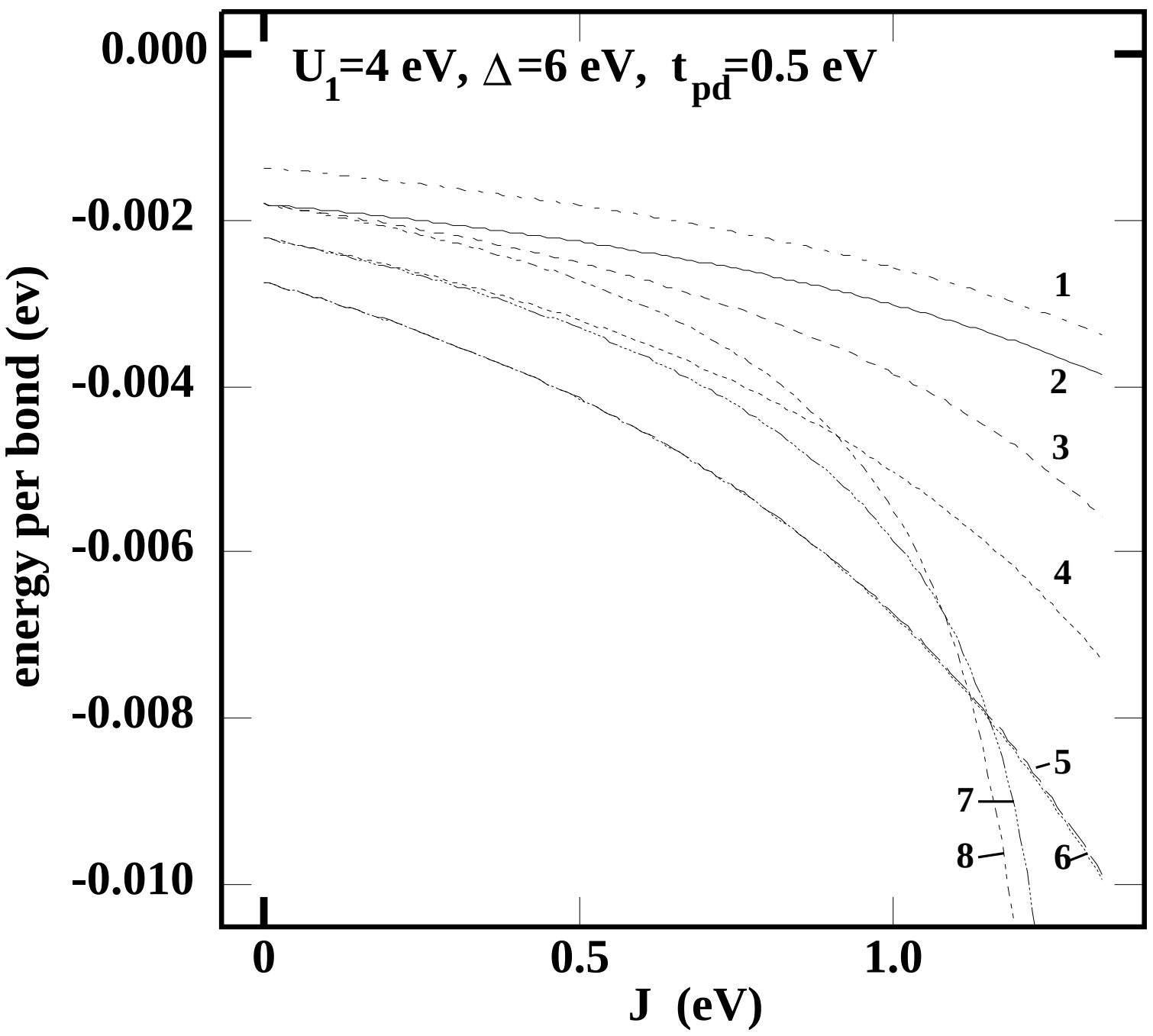

FIG. 16. Energy per bond for different configurations. (The configurations are shown in figure 15).

We used a realistic set of parameters ($U_1 = 4eV$, $\Delta = 6eV$ and $t_{pd} = 0.5eV$). The qualitative results, however, are not very sensitive on the actual choice of the numerical values. Notice that the energy difference between the lowest energy configurations extremely small (less than 5 mev). We checked that in the limit of extremely large $\Delta$ the energy gets minimal for ferromagnetic bonds for all values of $J$. In this case the ferromagnetism of a two band Hubbard model is recovered.

Clearly over a large range of the intraatomic exchange $J$ an almost non magnetic configuration, i. e. a configuration with a very small magnetic exchange constant, is lowest in energy. For extremely large $J$ the ferromagnetic configurations are favored while the antiferromagnetic contributions related to $J_1$ have always higher energies. For $LaTiO_3$, we conclude from our simple model that antiferromagnetism is strongly suppressed and the paramagnetic phase is stabilized justifying our use of the paramagnetic solution in our earlier descrciption of the experimental data in transition metal oxides using the one band Hubbard model.

It is clear that magnetic ordering takes place only below a quite low transition temperature $T_c$. Above $T_c$ the system is more or less frustrated. In this temperature regime the dynamical mean field theory has to be applied to a fully frustrated lattice to yield a reasonable description.

ACKNOWLEDGEMENTS: This work was supported by the National Science Foundation under grant DMR 95-29138. We would like to thank Dr. Natan Andrei, Dr. Herbert Neuberger, and Dr. Marcelo Rozenberg for useful discussions.

---